# Programmable spatiotemporal shaping for partially coherent structured wave packets

Zhao Zhang[1,2,*], Haihao Fan[3,*], Yaning Zhou[1,2], Weiran Li[1,2], Chaoliang Ding[4,5], Qiwen Zhan[6,†], Yangjian Cai[1,2,†], and Xin Liu[1,2,†]

[1]Shandong Provincial Key Laboratory of Light Field Manipulation Physics and Applications & School of Physics and Optoelectronics, Shandong Normal University, Jinan 250358, China;

[2]Collaborative Innovation Center of Light Manipulations and Applications, Shandong Normal University, Jinan 250358, China;

[3]School of Optical-Electrical and Computer Engineering, University of Shanghai for Science and Technology, Shanghai 200093, China;

[4]Tianjin Key Laboratory of Quantum Optics and Intelligent Photonics, School of Physical Science and Technology, Tianjin University of Technology, Tianjin 300384, China;

[5]State Key Laboratory of Photonics and Communications, Shanghai Jiao Tong University, Shanghai 200240, China;

[6]Zhejiang Key Laboratory of 3D Micro/Nano Fabrication and Characterization, Department of Electronic and Information Engineering, School of Engineering, Westlake University, Hangzhou 310030, China;

[*]These authors contributed equally to this work.

[†]Corresponding Authors: *zhanqiwen@westlake.edu.cn*; *yangjiancai@sdnu.edu.cn*; *xinliu@sdnu.edu.cn*

## Abstract

**Providing more coherent light with space-time coupling offers unprecedented opportunities across diverse scientific disciplines and areas of technology. However, in many applications, tailored laser beams with partial coherence in space and time are desired. Here, we demonstrate a programmable approach for simultaneously shaping the intensity and coherence distributions of spatiotemporally coupled partially coherent wave packets. By jointly engineering the fractal speckle of a dynamic scatterer and applying tailored modulation of the illumination function, we experimentally generate partially coherent laser beams with arbitrarily designed intensity and nontrivial coherence distributions. The proposed approach is compatible with ultrafast laser pulses, enabling the on-demand**

**generation of partially coherent space-time wave packets. The resulting spatiotemporally partially coherent wave packets exhibit non-separable space-time intensity profiles together with engineered spatiotemporal correlation distributions. This work provides a versatile framework for tailoring partially coherent light in space and time, with varied applications in relativistic laser-matter interaction, passive imaging and optical manipulation.**

## 1. INTRODUCTION

Structured light fields with non-separable spatiotemporal degrees of freedom exhibit complex forms in both space and time and have recently attracted considerable attention [1-3]. In this context, the ability to engineer optical wave packets with tailored spatiotemporal coupling on demand has emerged as a rapidly developing research direction, driven by their potential to uncover new physical phenomena and enable a wide range of applications. Representative advances in this area include the realization of spatiotemporal optical vortices carrying transverse orbital angular momentum [4-7], propagation-invariant space-time wave packets with controllable group velocities [8-11], and toroidal light vortices exhibiting torus-shaped energy flow distributions [12-15]. These advances in tailoring the deterministic spatiotemporal degrees of freedom of light have substantially broadened the scope of space-time coupling in optical topology [16-21], light-matter interactions [22,23], ultrafast light field characterization [24,25], as well as nonlinear optical processes [26,27].

Beyond the aforementioned first-order degrees of freedom of light, optical coherence represents a distinctive property of partially coherent beams, characterizing the statistical correlations of optical fields between pairs of space-time points [28]. Partially coherent beams play a pivotal role in applications involving spatiotemporally structured light, including passive imaging [29-31], optical cryptography [32-34] and metrology [35-38], and parallelized photonic computing [39-41], among other emerging areas in machining and lithography. Spatially, the coherence of a laser beam can be commonly tailored using a rotating diffuser based on the van Cittert-Zernike theorem [42,43]. Furthermore, a growing variety of three-dimensional structures, initially explored in coherent optical systems, have been successfully extended and observed in the statistical characteristics of partially coherent light [44]. However, current approaches for shaping spatiotemporally partially coherent beams remain largely confined to engineering their coherence structures, leaving their intensity distributions largely unexplored and typically restricted to space-time separable forms [45]. This limitation primarily arises from the challenge of manipulating temporal profiles of partially coherent space-time wave packets at ultrashort time scales. Nevertheless, no experimental platform has enabled the concurrent tailoring of the intensity and coherence properties of spatiotemporally partially coherent pulses.

In this paper, we present a general method for simultaneously shaping the intensity and coherence distributions of spatiotemporally coupled partially coherent light beams through the combined engineering of fractal speckle patterns of a dynamic scatterer and tailored modulation of the illumination function. This spatiotemporal shaping of both the intensity and coherence structures of the partially coherent laser beams is achieved in a programmable way. To verify the feasibility and versatility of the proposed approach, we first consider a spatially partially coherent beam and demonstrate the generation of a class of specifically correlated Schell-model beams with arbitrary intensity profiles. We further demonstrate that the proposed scheme enables the realization of statistically stationary optical wave packets with precisely engineered spatiotemporal intensity and coherence structures. Notably, the generated spatiotemporally partially coherent wave packets possess non-separable space-time characteristics, providing a flexible platform for exploring complex spatiotemporal coherence physics and advanced structured light applications.

## 2. ANALYTICS AND METHOD

According to coherence theory [28], the coherence properties of light are second-order statistical quantities characterized by the statistical correlations between stochastic optical fields evaluated at pairs of space-time points. In the space-time domain, a partially coherent beam can be described by the mutual intensity function associated with an ensemble of scalar random fields $\{U(\vec{r},\tau)\}$, which is defined as

$$\begin{aligned} G(\vec{r}_1,\vec{r}_2;\tau_1,\tau_2) &= \left\langle U(\vec{r}_1,\tau_1)U^*(\vec{r}_2,\tau_2)\right\rangle \\ &= A(\vec{r}_1,\tau_1)A^*(\vec{r}_2,\tau_2)u(\vec{r}_1,\vec{r}_2;\tau_1,\tau_2), \end{aligned} \tag{1}$$

where $\langle\cdot\rangle$ denotes ensemble averaging and $(\vec{r}_1,\vec{r}_2)$ and $(\tau_1,\tau_2)$ are arbitrary space and local time points. $u(\vec{r}_1,\vec{r}_2;\tau_1,\tau_2)$, and $|A(\vec{r},\tau)|^2$ for $\vec{r}_1=\vec{r}_2$, $\tau_1=\tau_2$ represent the coherence function and intensity of a partially coherent beam, respectively. To represent a physically realizable partially coherent light field [46], the mutual intensity function must satisfy the condition of non-negative definiteness and can therefore be expressed in the form

$$G(\vec{r}_1,\vec{r}_2;\tau_1,\tau_2)=\int I^2(\vec{v},\varpi)H(\vec{r}_1,\tau_1;\vec{v},\varpi)H^*(\vec{r}_2,\tau_2;\vec{v},\varpi)\mathrm{d}^2\vec{v}d\varpi, \tag{2}$$

where $I^2(\vec{v})$ is a non-negative function in mathematics and $\varpi=\omega-\omega_0$ represents the frequency offset relative to the carrier frequency $\omega_0$. $H(\vec{r},\tau;\vec{v},\varpi)$ denotes a propagation kernel function in the space-time domain, which is given by [45]

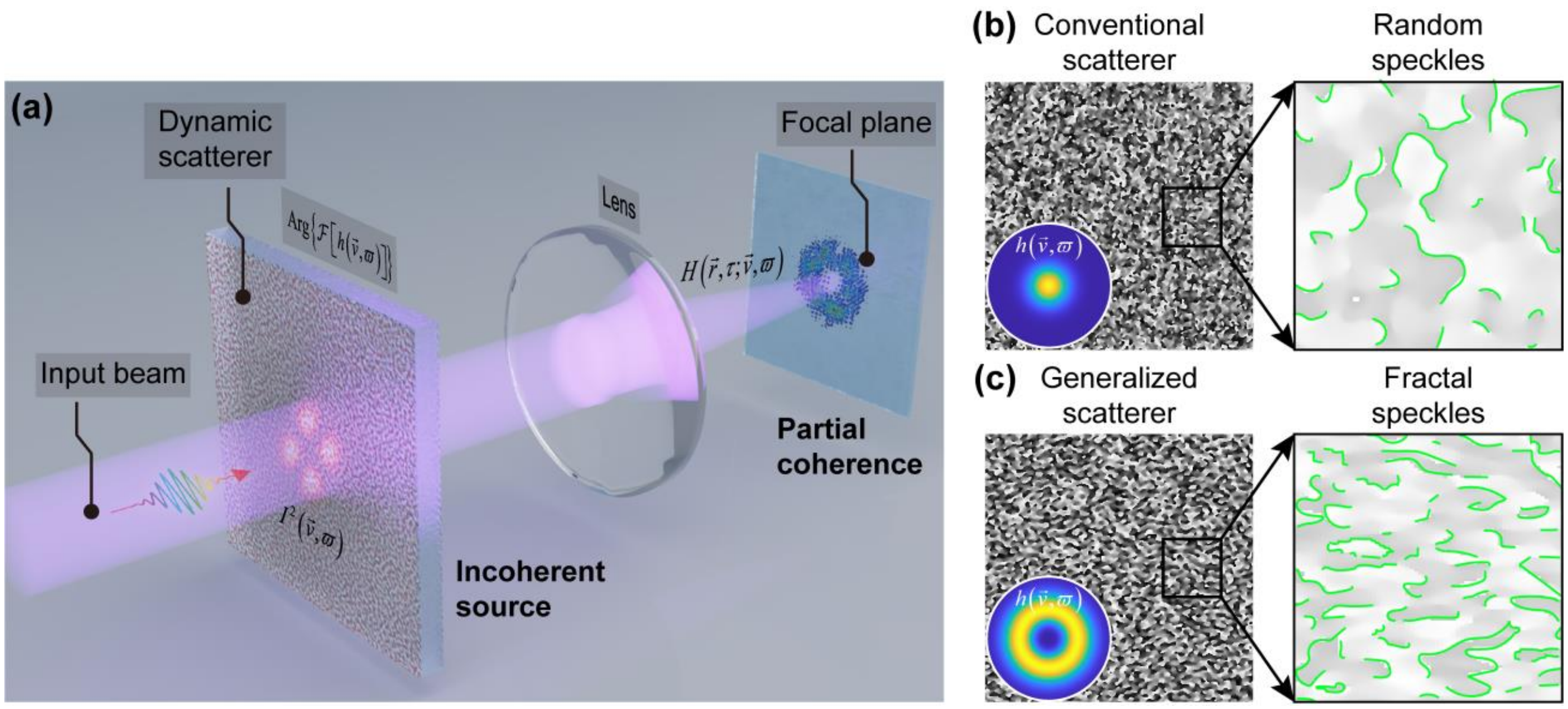

**Fig. 1**. (**a**) Conceptual schematic of the system required for shaping partially coherent laser beams with programmable intensity and coherence distributions. A coherent input laser beam illuminates a predesigned dynamic scatterer, generating an effective incoherent source with a specified intensity distribution. Following free-space propagation or linear dispersive evolution, the incoherent source transforms into the desired partially coherent beam in the far field (or equivalently, at the focal plane of a Fourier lens), exhibiting prescribed intensity profiles and tailored coherence structures. (**b**) Conventional scatterer producing random speckles characterized by a Gaussian-type $h(\vec{v},\varpi)$ [insets show the associated $h(\vec{v},\varpi)$]. (**c**) Generalized scatterer producing fractal speckles characterized by a non-Gaussian-type $h(\vec{v},\varpi)$.

$$\begin{aligned}H(\vec{r},\tau;\vec{v},\varpi)=&(k_0/2\pi iz)\exp\left[ik_0\left(|\vec{r}|^2+|\vec{v}|^2\right)/2z\right]\\&\exp\left(i\beta_2\varpi^2z/2-ik_0\vec{r}\cdot\vec{v}/z-i\varpi\tau\right),\end{aligned} \tag{3}$$

where $k_0=2\pi/\lambda_0$ is a wavenumber and $\beta_2$ is a group velocity dispersion coefficient. Figure

1(a) illustrates a representative configuration for generating partially spatially coherent beams from an incoherent pseudo-thermal source, wherein a dynamic scatterer (such as a rotating diffuser) introduces random phase fluctuations across the incident beam, thereby producing a spatially incoherent scattered field [$I^2(\vec{v},\varpi)$ is the incoherent intensity distribution] immediately after the scatterer. This incoherence condition is satisfied when the beam spot size on the scatterer, $I^2(\vec{v},\varpi)$, is larger than the characteristic inhomogeneity scale of the mosaic speckle structure of the scatterer [42]. Notably, the characteristic of a conventional phase-only scatterer, such as the ground glass plate, is governed by its power spectral density $h(\vec{v},\varpi)$, which typically exhibits a finite Gaussian profile, as illustrated in Fig. 1(b). In the present work, we instead consider an arbitrary distribution for $h(\vec{v},\varpi)$, as shown in Fig. 1(c). By leaving out the wavefront curvature and considering the Fourier-transform representation in Eq. (3), the corresponding phase map of the scatterer is described through $\mathrm{Arg}\{\mathcal{F}[h(\vec{v},\varpi)\times\Re(\vec{v},\varpi)]\}$, where $\Re(\vec{v},\varpi)$ denotes a complex random function and obeys a Gaussian statistic, and $\mathcal{F}$ represents the Fourier transform. The resulting phase shows fractal speckles and is uniformly distributed over the interval $(0,2\pi]$. The speckle patterns have a definite speckle size which is inversely proportional to a linear extent of $h(\vec{v},\varpi)$ while the morphology of the speckled fractal pattern is determined by the shape of $h(\vec{v},\varpi)$ [47]. The physical correspondence between the scatterer microstructure and the far-field intensity is established by the power spectral density $h(\vec{v},\varpi)$. The spectrum specifies the mean weights of the scattered spatial-frequency components, and the Fourier lens maps these components to focal-plane positions. When the illumination spot $I^2(\vec{v},\varpi)$ is sufficiently broad, this convolution kernel becomes narrow, and the ensemble-averaged far-field intensity approximately reproduces the prescribed spectrum [Supplement]. As compared in Figs. 1(b) and 1(c), although the resulting mosaic scatterers appear visually similar at the macroscopic level, their microscopic speckle structures—particularly the distributions of random phase dislocations—exhibit distinct fractal characteristics [47]. We have now unlocked dual degrees of freedom, $I(\vec{v},\varpi)$ and $h(\vec{v},\varpi)$, at the modulation plane for sculpting the target beams. In the following, we separately verify the shaping performance in the spatial and temporal domains.

## 3. EXPERIMENT AND MODELS

The key idea of the present paper is straightforward and intuitive: with a custom-designed dynamic scattering medium (replacing the conventional ground-glass diffuser) and its tailored illumination function, the intensity and coherence structures of the generated partially coherent light at the focal region can be controlled simply. Motivated by the two-dimensional space-time duality, which establishes a fundamental correspondence between spatially structured light and spatiotemporal beams via paraxial diffraction and narrowband dispersion described by Eq. (3), we begin by experimentally verifying the proposed scheme through shaping the intensity distribution and coherence properties of spatially partially coherent light beams. Specifically, a phase-only spatial light modulator (SLM) displaying dynamically refreshed random holograms was employed to emulate a rotating diffuser, as illustrated in Fig. 2(a). The incident laser beam with an intensity profile $I^2(\vec{v})$ was first scattered by the dynamic diffuser and then focused onto a camera through a Fourier lens. By continuously refreshing the random holograms displayed on the SLM, the dynamic scatterer generated a new speckle realization at the focal plane approximately every 200 ms. Consequently, the synthesized partially coherent beam was characterized by evaluating its time-averaged intensity distribution and the squared modulus of the complex degree of coherence, which can be respectively expressed as

$$S(\vec{r},\tau)=\frac{1}{N}\sum_{n=1}^{N}I^{S}(\vec{r},\tau), \tag{4}$$

and

$$\left|u(\vec{r}_1,\vec{r}_2;\tau_1,\tau_2)\right|^2=\frac{\left\langle I^S(\vec{r}_1,\tau_1)I^S(\vec{r}_2,\tau_2)\right\rangle}{\left\langle I^S(\vec{r}_1,\tau_1)\right\rangle\left\langle I^S(\vec{r}_2,\tau_2)\right\rangle}-1, \tag{5}$$

where $I^S(\vec{r},\tau)$ is the instantaneous speckled intensity and *N* is the number of the ensemble to be averaged.

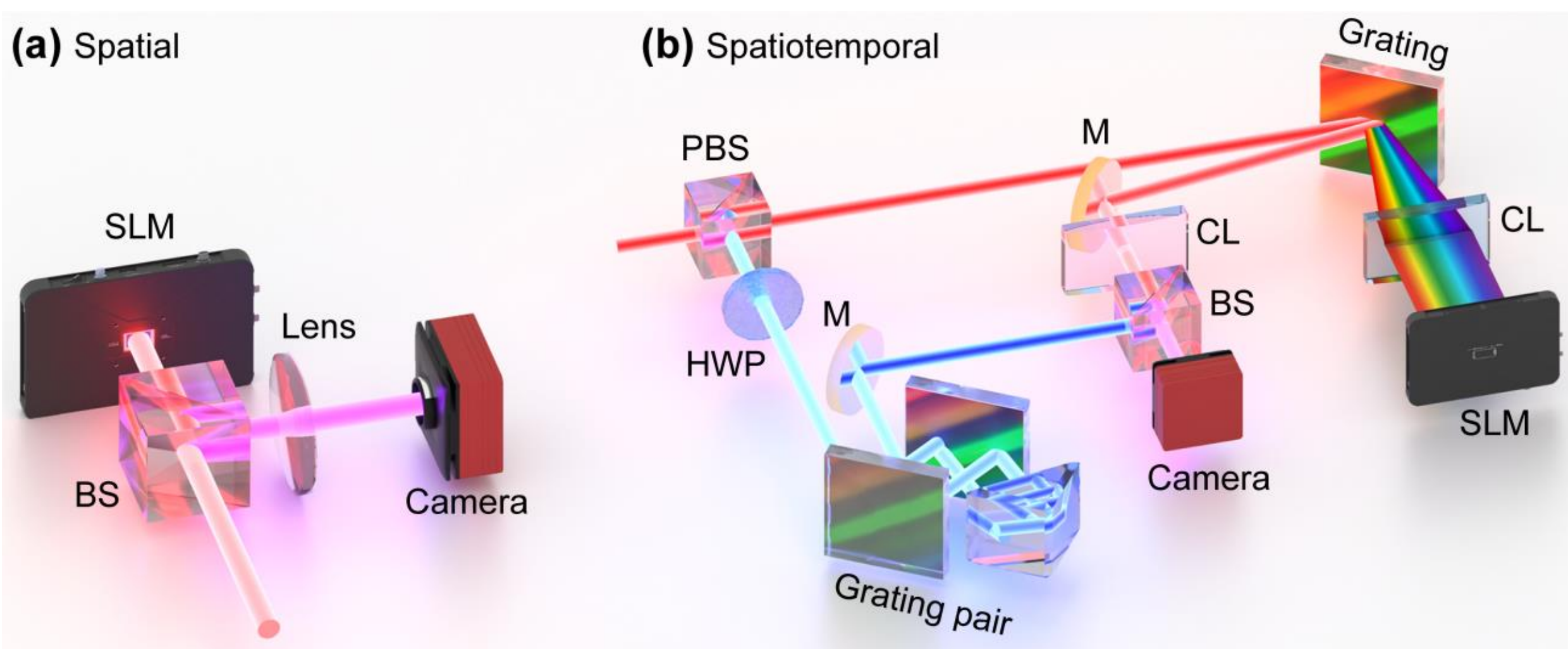

**Fig. 2**. Experimental setups for demonstrating the shaping of partially coherent light beams in the (**a**) spatial domain and (**b**) spatiotemporal domains. SLM, phase-only spatial light modulator; BS, beam splitter; PBS, polarization beam splitter; CL, cylindrical lens; HWP, half-wave plate.

Figure 3 illustrates the generation of spatially partially coherent beams with donut-shaped intensity profiles and different correlation structures using the proposed approach. Without lack of generality, in the experiments and simulations, the $h(\vec{v})$ function of the scatterer is set as $|\vec{v}|^2/w_0^2 \exp(-|\vec{v}|^2/w_0^2)$ with $w_0 = 2$ mm. The target degree of coherence functions are selected as follows

(1) Hermite-Gaussian correlation function of $m,n$-th (HGC$_{m,n}$) [48]:

$$\mu(\vec{r}_1,\vec{r}_2)=\frac{H_{2m}\left[\left(x_2-x_1\right)/\sqrt{2}\sigma_x\right]H_{2n}\left[\left(y_2-y_1\right)/\sqrt{2}\sigma_y\right]}{H_{2m}(0)H_{2n}(0)}\exp\left(-\frac{|x_2-x_1|^2}{2\sigma_x^2}-\frac{|y_2-y_1|^2}{2\sigma_y^2}\right), \tag{6}$$

where $H_m[\cdot]$ denotes the Hermite polynomial of the order $m$. $\sigma_x$ and $\sigma_y$ being the transverse coherence widths along the *x* and *y* directions.

(2) Laguerre-Gaussian correlation function of $n$-th (LGC$_n$) [49]:

$$\mu(\vec{r}_1,\vec{r}_2)=L_n\left(\frac{|\vec{r}_1-\vec{r}_2|^2}{2\sigma^2}\right)\exp\left(-\frac{|\vec{r}_1-\vec{r}_2|^2}{2\sigma^2}\right), \tag{7}$$

where $L_n(\cdot)$ denotes the Laguerre polynomial of the order $n$. $\sigma$ being the transverse coherence width.

## 4. SPATIAL SHAPING

Figure 3(a) presents the experimental and simulated time-averaged intensity distributions of the generated partially coherent beams with nonconventional correlation functions. As expected, the resulting intensity profiles exhibit the desired donut-shaped distribution, which is determined by the power spectral density function $h(\vec{v})$ of the dynamic scatterer. The scripts used to perform the simulations in this work are available at [50]. Figures 3(b)-3(d) and 3(e)-3(g) present the experimental and simulated results for the squared modulus of the degree of coherence, $|u(\vec{r}_1,\vec{r}_2)|^2$, of the generated specially correlated partially coherent beams with different spatial coherence structures. The insets show representative instantaneous speckle intensity patterns of the corresponding partially coherent laser beams. The coherence structures $|u(\vec{r}_1,\vec{r}_2)|^2$, evaluated at the reference point indicated in Fig. 3(a) using 1500 speckle realizations, exhibit well-defined $HGC_{1,1}$, $HGC_{1,1}$ and $LGC_2$ distributions. Overall, the experimentally measured intensity profiles and coherence distributions of the generated partially spatially coherent beams show excellent agreement with the numerical simulations.

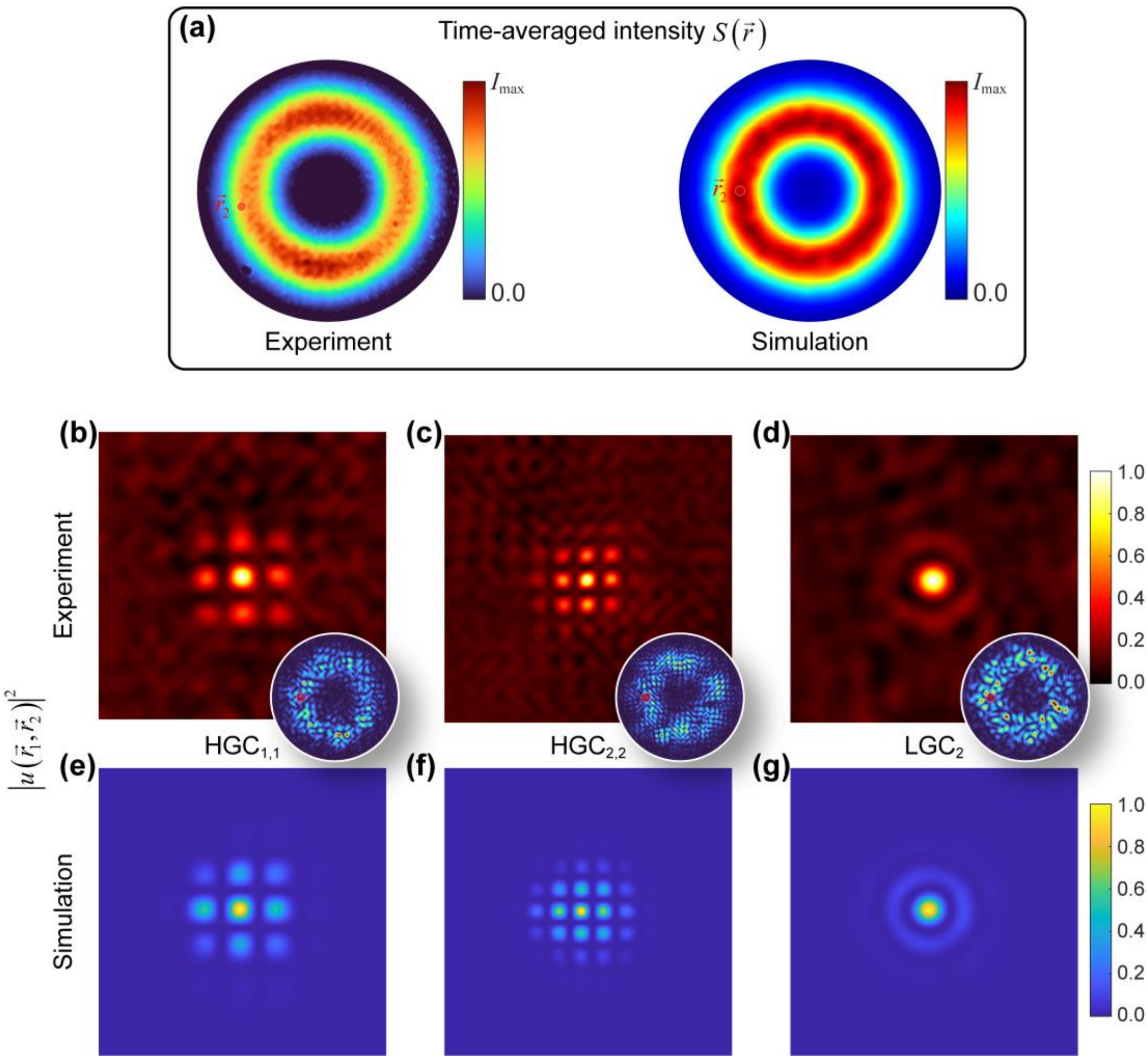

**Fig. 3**. (**a**) Time-averaged intensity distributions of the experimentally generated (left) and numerically simulated (right) spatial partially coherent beams with donut-shaped intensity profiles. (**b**)-(**g**) Measured (top) and simulated (bottom) $|\mu(\vec{r}_1,\vec{r}_2)|^2$ at the reference point indicated in (a) with (**b**),(**e**) Hermite-Gaussian correlation $HGC_{1,1}$, (**c**),(**d**) Hermite-Gaussian correlation $HGC_{2,2}$ and (**d**),(**g**) Laguerre-Gaussian correlation $LGC_2$. The insets show the experimentally recorded instantaneous intensity distributions of the generated partially coherent beams. The results are obtained from averaging over 1500 speckled realizations.

The proposed scheme enables flexible tailoring of the intensity distribution of specially correlated Schell-model beams. To demonstrate this capability, we generated a spatially partially coherent beam with programmable transverse intensity and coherence structures in the focal region, as shown in Fig. 4. In particular, for applications such as laser heat treatment and surface annealing in microelectronics, beam shaping with both uniform energy fluence distribution and low spatial coherence is highly desirable. Therefore, a flat-top HGC beam was generated, as illustrated in Fig. 4(a), producing a focal intensity distribution with a uniform flat-top profile and sharply defined edges. The spatial correlation structures evaluated at different reference points are presented in Figs. 4(b)-4(d), demonstrating the reference-independent Hermite-Gaussian correlated Schell-model characteristics of the generated beam. In addition, compared with the focal intensity distribution, Figs. 4(b)-4(d) further reveal the low-spatial-coherence (around 300 μm) feature of the target beam. Figures 4(e)-4(h) present the experimentally generated partially coherent beams with predesigned complex intensity patterns and coherence structures. Such beams, engineered to exhibit tailored intensity distributions and low spatial coherence simultaneously, are of significant importance in applications such as optical material processing and three-dimensional printing, where enhanced uniformity and reduced coherence-induced interference effects are critical.

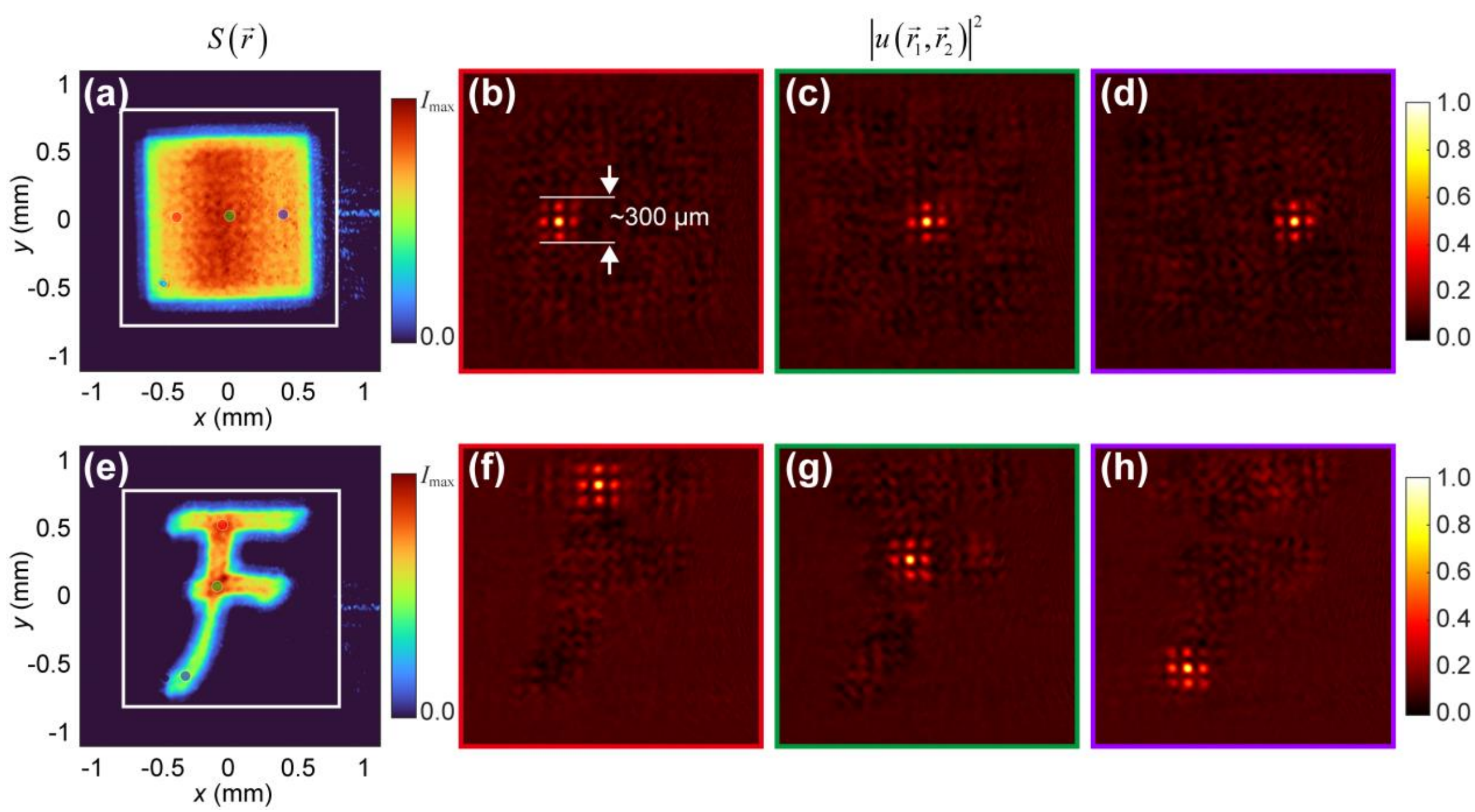

**Fig. 4**. (**a**) Measured intensity distribution of a spatial Hermite-Gaussian correlated beam with a flat-top intensity profile. (**b**)-(**d**) Measured $|\mu(\vec{r}_1,\vec{r}_2)|^2$ at different reference points indicated in (a). (**e**) Measured intensity distribution of a spatial Hermite-Gaussian correlated beam with an arbitrary intensity profile. (**f**)-(**h**) Measured $|\mu(\vec{r}_1,\vec{r}_2)|^2$ at different reference points indicated in (e). The results are obtained from averaging over 1500 speckled realizations.

## 5. SPATIOTEMPORAL SHAPING

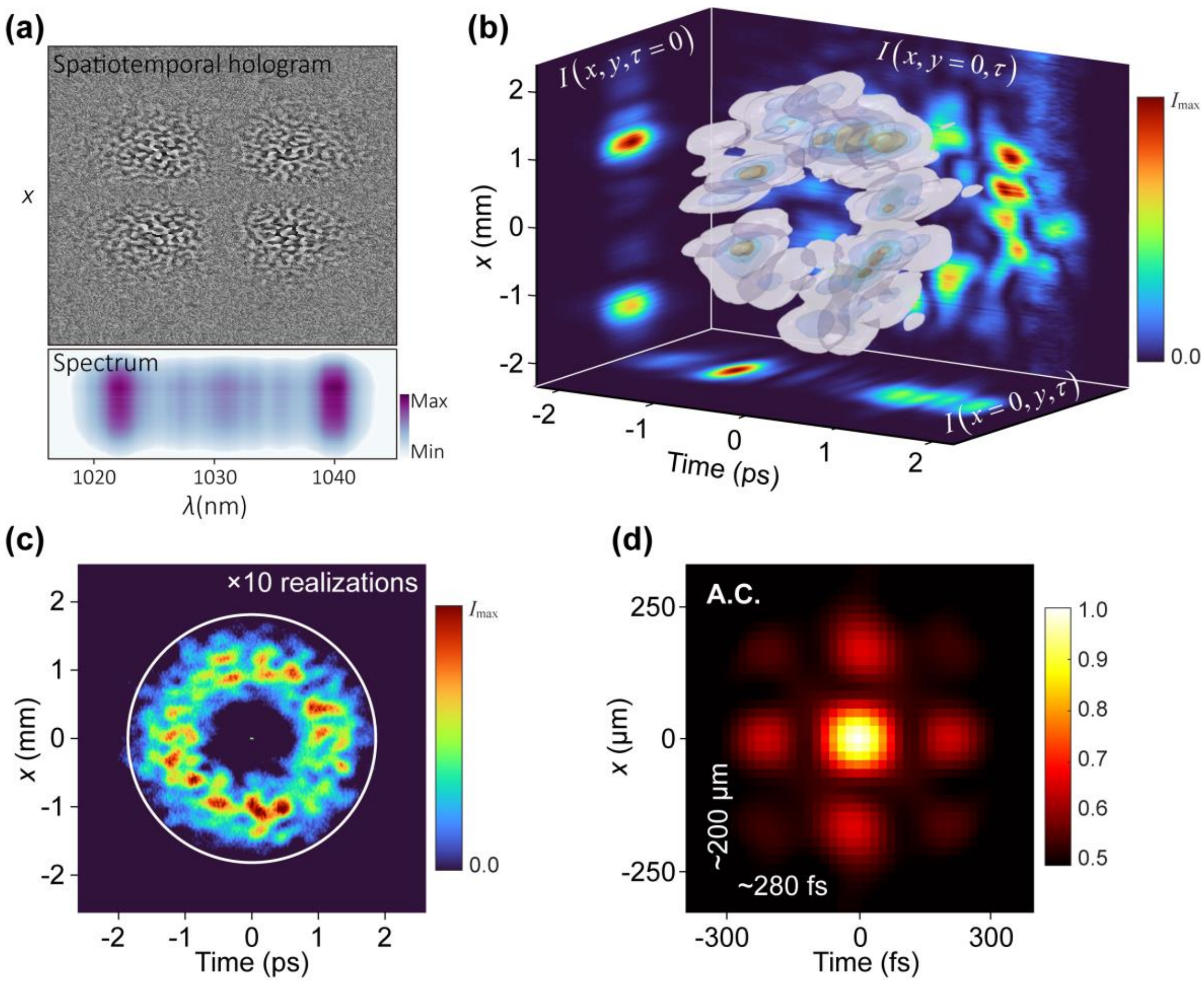

**Fig. 5**. (**a**) Random phase mask designed for generating partially coherent space-time wave packets with a Hermite-Gaussian correlation structure and a non-separable donut-shaped intensity profile. Bottom: spectrum distribution at the SLM plane. (**b**) 3D intensity distribution of a single-shot output pulse. (**c**) Spatiotemporal intensity distribution obtained by averaging over 10 speckled realizations, exhibiting a space-time non-separable donut-shaped profile. (**d**) Measured autocorrelation distribution corresponding to the single-shot output pulse shown in (b).

Beyond spatial-only shaping of partially coherent beams, the proposed scheme can be readily reconfigured to tailor both the spatiotemporal intensity and coherence properties of partially coherent wave packets on demand. Figure 2(b) illustrates the experimental setup for generating and characterizing partially spatiotemporally coherent pulsed beams, which consists of a two-dimensional pulse shaper and a Mach-Zehnder-type scanning interferometer [45]. Briefly, a pulsed beam with a bandwidth of approximately 20 nm centered at 1030 nm, emitted from a mode-locked fiber laser, is first split into two coherent replicas—an object pulse (red) and a probe pulse (blue)—using a polarizing beam splitter (PBS). The spectrum of the object pulse is subsequently dispersed by a grating and collimated by a cylindrical lens (CL) before being projected onto a phase-only spatial light modulator (SLM, Holoeye GAEA-2, 3840×2160 pixels with a 3.74 μm pitch and 60 fps). The SLM emulates a rotating diffuser by scattering the incident spatial spectrum through complex-amplitude holograms [6]. A complete statistical ensemble is generated by sequentially refreshing the random phase distributions encoded on the holograms. The scattered spectrum is then recombined using a second cylindrical lens and grating to reconstruct the pulse through temporal Fourier transformation, while an additional lens performs the spatial Fourier transformation. At each realization, the generated field can be regarded as an individual coherent mode whose spatiotemporal profile is fully characterized using a Mach–Zehnder-type scanning interferometer. Specifically, the probe pulse is directed through a pulse compressor composed of a parallel-grating pair and a right-angle prism, producing a nearly Fourier-transform-limited pulse with a duration of approximately 120 fs, substantially shorter than that of the object wave packet. The compressed probe pulse subsequently interferes with the object pulse at a selected temporal slice, generating interference fringes that encode the local spatiotemporal information. By scanning the relative delay between the two arms, the complete spatiotemporal profile of each ensemble realization can be computationally reconstructed from the delay-dependent interference patterns. It should be noted that the data acquisition processes for spatial-domain and spatiotemporal-domain measurements differ. A spatial realization is recorded directly as an intensity image, whereas one spatiotemporal realization is reconstructed from a complete delay scan acquired using a fixed hologram. Such scans considerably increase the per-realization acquisition time, which is why different ensemble sizes are employed.

Figure 5 presents the experimental results demonstrating the generation of a partially spatiotemporally coherent wave packet with space-time non-separability. The applied spatiotemporal hologram (a representative random phase mask), shown in Fig. 5(a), was loaded onto the SLM to modulate the spatial-spectral wavefront (bottom) of the incident chirped pulse. The experimental parameters were chosen to be the same as those used in Figs. 2(a) and 2(b). Figure 5(b) presents the experimentally measured three-dimensional instantaneous intensity distribution of the generated partially spatiotemporally coherent pulsed beam. It can be clearly observed that localized spatial and temporal speckle structures are confined within a donut-shaped region. After ensemble averaging over 10 speckled realizations of the resulting partially coherent space-time wave packets, the intensity distribution exhibits a clear space-time non-separable donut-shaped profile, as shown in Fig. 5(c). The spatiotemporal coherence structure can be conveniently characterized through the autocorrelation of the measured speckle pattern. As shown in Fig. 5(d), the experimentally measured spatiotemporal coherence distribution exhibits a non-Gaussian profile with the desired $HGC_{1,1}$ structure. Through theoretical fitting, the spatial and temporal coherence widths are estimated to be approximately $\sigma_x = 200$ um and $\sigma_\tau = 280$ fs, respectively.

## 6. DISCUSSION AND CONCLUSION

In summary, we propose an experimental approach for simultaneously tailoring the intensity and coherence properties of partially coherent beams in the spatial and spatiotemporal domains. Unlike conventional diffusers, such as ground-glass plates, the dynamic scatterer employed here consists of a random phase mask whose microstructures are engineered to exhibit a prescribed fractal pattern. The incoherent source generated behind the diffuser evolves, upon propagation under free-space or linear dispersion, into a partially coherent beam with prescribed intensity and statistical correlation distributions. In this framework, the beam intensity profile and coherence structure are independently controlled in a programmable way by the fractal microstructure of the diffuser and the illumination distribution, respectively. Partially spatially coherent beams with arbitrary intensity distributions and prescribed spatial correlations are experimentally realized and shown to conform to the Schell-model beam. By further integrating this scheme with the spatiotemporal holographic pulse shaper, we achieve, for the first time, the controlled generation of partially coherent wave packets with tailored intensity and coherence structures in a non-separable space-time domain. This advancement is particularly significant, as simultaneous and independent control of both intensity and coherence in the spatiotemporal domain, in a fully flexible and programmable manner, is an ongoing issue. We anticipate that the ability to fully engineer partially coherent wave packets will open new opportunities in ultrafast structured-light science, including light-matter interactions, ultrafast light transport in complex media, and coherent control in quantum systems.

The proposed experimental approach is compact and cost-effective because it only requires a single dynamic phase-only diffuser for spatiotemporal shaping of an input beam. Last and notable, the transverse spot size of the illumination incident on the scatterer must be larger than the characteristic inhomogeneity scale of the scatterer [Supplement]. This condition is essential to ensure the formation of an effectively incoherent source and to justify the applicability of the van Cittert-Zernike framework [42,45]. Furthermore, the dynamic diffuser used here is fundamentally phase-only and isotropic. Such a scatterer can be implemented using diffractive optical element (DOE) fabrication techniques and rendered dynamic by varying the relative position between the scattering medium and the illumination [51-53].

**Funding.** National Natural Science Foundation of China (W2441005 [Y.C.], 12192254 [Y.C.], 12534014 [Y.C.], 12434012 [Q.Z.], 62535013 [Q.Z.], 12604505 [X.L.], 12547149 [X.L.]), National Key Research and Development Program of China (2022YFA1404800 [Y.C.]), Natural Science Foundation of Shandong Province (ZR2025ZD21 [Y.C.]), Key Research and Development Program of Shandong Province (2024JMRH0105 [Y.C.]), State Key Laboratory of Photonics and Communications of China (2026QZKF020 [C.D.]). X.L. acknowledges support from the Young Talent of Lifting engineering for Science and Technology in Shandong (SDAST2026 QTA028) and the China Postdoctoral Science Foundation (2026M793716).

**Disclosures.** X.L., Z.Z. and Y.C. are authors on a patent application related to this work filed by Shandong Normal University. The authors declare no competing interests.

**Data availability.** The data that support the findings of this study are available from the corresponding authors upon request.

**Author contributions.** X.L. conceived the idea and directed the project; Z.Z. and H.F. conducted the experiments and measurements; Y.Z., W.L., C.D. and X.L. performed the simulations and the analysis; X.L. wrote the paper with input from Z.Z and H.F.; Q.Z., Y.C. and X.L. co-supervised the project. All authors contributed to the discussion and revision of the manuscript.

**Supplemental document.** See Supplement for supporting content.